\documentclass[fleqn,twoside]{article}
\usepackage{espcrc2}
\usepackage{amssymb}
\usepackage{color}

\usepackage{graphicx}
\usepackage[figuresright]{rotating}

\newcommand{\bm}[1]{{\mbox{\boldmath$#1$}}}
\newcommand{\case}[2]{{\textstyle\frac{#1}{#2}}}
\newcommand{\Dirac}{{\kern+0.1em /\kern-0.65em D}}
\newcommand{\LEL}{LE$\mathcal{L}$}
\newcommand{\LQCD}{\ensuremath{\Lambda_{\rm QCD}}}
\newcommand{\tr}{\mathop{\rm tr}}

\newcommand{\nrvel}{{\textsl{v}}}
\newcommand{\vlabel}{{v}}

\title{Heavy Quarks and Lattice QCD
\hfill {\normalsize FERMILAB-Conf-03/366-T}}

\author{Andreas S. Kronfeld\address[FNAL]{Theoretical Physics Department, 
	Fermi National Accelerator Laboratory, 
	Batavia, Illinois 60510, USA}
	\hfill hep-lat/0310063}
       
\begin{document}

\begin{abstract}
	This paper is a review of heavy quarks in lattice gauge theory,
	focusing on methodology.
	It includes a status report on some of the calculations that are
	relevant to heavy-quark spectroscopy and to flavor physics.
\vspace{-10pt}
\end{abstract}

\maketitle

\section{MOTIVATION}
\label{sec:intro}

\begin{figure}[b!]
	\vspace*{-19pt}
	\centering
    \includegraphics[width=72mm]{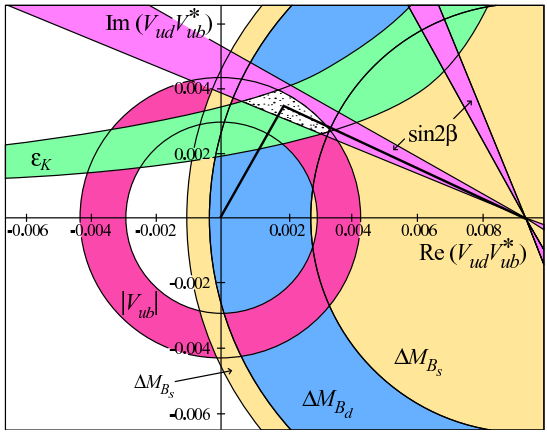}
	\vspace*{-24pt}
    \caption[fig:ut:pdg]{Unitarity triangle from Ref.~\cite{Hagiwara:fs}.}
    \label{fig:ut:pdg}
\end{figure}

The study of flavor- and $CP$-violation is a vital part of particle
physics~\cite{Hazumi:2003bl}.
Often lattice QCD is needed to connect experimental measurements
to the fundamental couplings of quarks, which, in the Standard Model, 
are elements of the Cabibbo-Kobayashi-Maskawa (CKM) matrix.
Usually the test of CKM is drawn as a set of constraints on the apex of
the so-called unitarity triangle (UT).
The Particle Data Group's version~\cite{Hagiwara:fs}
is shown in Fig.~\ref{fig:ut:pdg}.
Apart from the wedge $\sin2\beta$, theoretical uncertainties dominate,
and everyone wonders whether they are reducible and, if so, reliable.
Indeed, as Martin Beneke~\cite{Beneke:2002ks} put it at Lattice 2001, 
the ``Standard UT fit is now entirely in the hands of Lattice QCD
(up to, perhaps, $|V_{ub}|$).''

The needed hadronic matrix elements are among the simplest in
lattice~QCD, so we can hope to carry out a full and reliable error
analysis.
Two criteria are key:
First, there must be one stable (or very narrow) hadron in the
initial state and one or none in the final state; 
second, the chiral extrapolation must be under control.
Such quantities can be called gold-plated, to remind us that
they are the most robust.
(They are conceptually and technically much simpler than non-leptonic
decays, or resonance masses and widths.)
Moreover, realistic, unquenched simulations for gold-plated
quantities now seem to be feasible~\cite{Davies:2003ik,Gottlieb:2003lt}.

\begin{figure}[b!]
	\centering
	\vspace*{-14pt}
    \includegraphics[width=75mm]{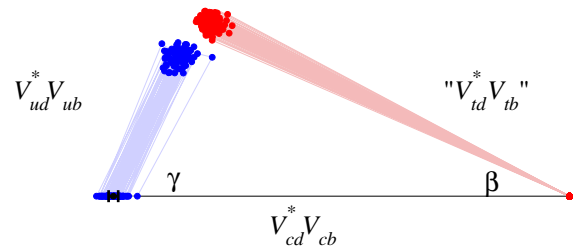}
	\vspace*{-34pt}
    \caption[fig:ut:reloaded]{A possible unitarity triangle in
    a few years. 
	The scatter covers approximately~$5\sigma$.}
    \label{fig:ut:reloaded}
\end{figure}

Much is at stake.
Fits to the CKM matrix are often described as over-constraining the
Standard Model, but they are really a test.
With (reliable) error bars of a few percent one can 
imagine a picture like Fig.~\ref{fig:ut:reloaded}.
The base, the angle $\gamma$, and the left side come from $B$ (and other)
decays that proceed at the tree level of the electroweak interactions.
They can be considered to measure the UT's apex.
The side labeled ``$V_{td}^*V_{tb}$'' is obtained from the
frequency of $B^0$-$\bar{B}^0$ mixing; the angle $\beta$ through
the interference of decay with and without mixing.
Standard mixing proceeds through box diagrams, so
non-Standard processes could compete.
Thus, this side tests the CKM theory.
It could fail the test by missing the apex, thereby inspiring a grand
revision of the program of flavor physics: ``The CKM Matrix Reloaded.''

Further motivation for lattice QCD with heavy quarks has 
emerged this year.
The BaBar~\cite{Aubert:2003fg}, CLEO~\cite{Besson:2003cp} and 
Belle~\cite{Abe:2003jk} experiments have observed the
lowest-lying $J^P=0^+$ and $1^+$ states in the $D_s$~system.
They lie slightly below the $D^{(*)}K$ thresholds, so they are narrow,
decaying via isospin violation to $D_s^{(*)}\pi$.
Several groups have started to calculate the masses with lattice QCD;
see Sec.~\ref{subsec:Ds}.
The SELEX experiment reports not one or two but five baryons 
that presumably contain two charmed quarks~\cite{Mattson:2002vu}.
Corroboration from lattice QCD would be useful.
A~pilot study appeared during the conference~\cite{Flynn:2003vz}.

The rest of this paper is organized as follows.
In Sec.~\ref{sec:methods} the problem of heavy quarks in lattice gauge
theory is deconstructed into three elements: a discretization of the
heavy quark(s), an effective field theory for describing cutoff effects,
and a set of calculations of the short-distance behavior for matching
to~QCD.
The most widely used methods are then cast into this framework.
Sec.~\ref{sec:cutoff} reviews the effective field theories needed to 
tame heavy-quark discretization effects, leading up to a semi-quantitative 
comparison of the methods.
The issue of renormalon shadows is re-examined in Sec.~\ref{sec:renshad}.
A new technique is discussed in Sec.~\ref{sec:new}.
Sec.~\ref{sec:results} discusses the mass spectrum and matrix elements
needed for the Standard CKM fit.
A~short summary and outlook are in Sec.~\ref{sec:summary}.

\section{METHODS FOR HEAVY QUARKS}
\label{sec:methods}

To connect lattice calculations of heavy-quark physics to experiment,
we must confront three principal concerns:
the quenched approximation, heavy-quark discretization 
effects, and the chiral extrapolation.
The quenched approximation is the most worrisome to experimenters and
phenomenologists, because its error is unquantifiable.
Fortunately, it seems to be going away (at last).
With 3 improved staggered sea quarks we find agreement with 
experiment for a wide variety of gold-plated masses and decay 
constants~\cite{Davies:2003ik,Gottlieb:2003lt}.
Some issues should be clarified~\cite{Jansen:2003lt}, 
but it is clearly time to face the other two problems.

Heavy-quark discretization effects are vexing because, with available
computing resources, the bottom quark's mass in lattice units is not
small, $m_ba\not\ll1$.
(The charmed quark's mass is also not very small.)
When $ma\ll1$, as for light quarks, one can control and quantify cutoff
effects with a Symanzik effective field theory~\cite{Symanzik:1979ph}.
This kind of idea can be extended to heavy 
quarks~\cite{Kronfeld:2000ck,Harada:2001fi}.
The ensuing insights then allow 
us to compare and contrast the size and nature of heavy-quark 
discretization effects in the various techniques.

The chiral extrapolation arises, as elsewhere in lattice QCD, because
quark masses as small as those of the up and down quarks are not feasible
computationally.
The key issue~\cite{Bernard:2002yk} is whether the numerical data exhibit
the curvature characteristic of chiral perturbation theory ($\chi$PT).
If so, we can trust $\chi$PT to guide the extrapolation.
If not, the extrapolation is still necessary, but its uncertainty becomes
difficult to estimate.
Indeed, in at least one important example this uncertainty has
been underestimated~\cite{Kronfeld:2002ab}.

There are many ways to treat heavy quarks in lattice gauge theory.
It is instructive to deconstruct them into three essential elements, 
illustrated in Table~\ref{tbl:methods}.
\begin{table*}[t!p]
	\caption{The main elements for incorporating quarks with mass
	$m_Q\gg\LQCD$ into lattice gauge theory.}
	\begin{tabular*}{\textwidth}{l@{\extracolsep{\fill}}l@{\extracolsep{\fill}}l}
	\hline
		heavy-quark discretization & effective field theory tools & 
		renormalization \& matching \\
		\hline
		(improved) Wilson quarks & Symanzik \LEL: for $m_Qa\ll1$ & 
			perturbative: tadpole tree-level\\
		 & 
		\hphantom{Symanzik \LEL: }for $m_Qa\not\ll1$ & 
		\hphantom{perturbative: }1- or 2-loop \\
		static quarks (+ insertions) & HQET (for $\bar{q}Q$) & 
			non-perturbative  \\
		non-relativistic quarks & NRQCD (for $\bar{Q}Q$) & combination: 
			$Z_A = \rho_A^{\mathrm{PT}}Z_V^{\mathrm{NP}}$  \\
	\hline
	\end{tabular*}
	\label{tbl:methods}
	\vspace{-12pt}
\end{table*}
The first is the choice of discretization---%
a lattice fermion field and action.
By ``Wilson quarks'' I mean four-component lattice fermion fields with 
any action using Wilson's solution of the doubling
problem~\cite{Wilson:1975hf}.
``Static quarks'' are essentially one-component~\cite{Eichten:1987xu}; 
spin-dependent effects are incorporated as
insertions~\cite{Eichten:1990vp}.
``Non-relativistic quarks'' have two components~\cite{Lepage:1987gg}, 
and the action is a discretization of a non-relativistic field
theory~\cite{Caswell:1986ui}.
Lattice gauge theories with any of these quark fields are compatible with
the heavy-quark symmetries that emerge in
QCD~\cite{Shifman:1987rj,Isgur:1989vq}.

%
To run numerical simulations one needs only the discretization.
To interpret the simulation data, however, one needs more.
A~typical simulation has unphysical values of some of the scales, such 
as lattice spacing $a\neq0$ and light quark mass much larger than the 
down quark's mass.
A theoretical framework is needed to get from simulations of lattice
gauge theory (LGT) to the target quantum chromodynamics (QCD).
A~powerful set of frameworks is based on effective field theories.
Indeed, most theoretical uncertainties in lattice QCD can (and should) be
controlled and quantified with effective field
theories~\cite{Kronfeld:2002pi}.

In the context of discretization effects, the best-known effective field
theory is Symanzik's theory of cutoff effects~\cite{Symanzik:1979ph}.
The central concept is a local effective Lagrangian (\LEL) of a 
continuum field theory.
It is also possible, when a quark is heavy, to use the (continuum)
heavy-quark theories HQET (heavy-quark effective theory) and NRQCD
(non-relativistic QCD)~\cite{Kronfeld:2000ck,Harada:2001fi}.
The next section discusses these ideas in detail.
Here I note only that I distinguish HQET from NRQCD by the way they 
classify interactions, as dictated by the physics
of heavy-light and heavy-heavy hadrons, respectively.

Effective field theories separate dynamics at short distances 
($a$, $1/m_Q$) from those at long distances 
(1/$\LQCD$, $1/m_q$, $L$).
To renormalize LGT (or to ``match'' LGT to QCD), one must calculate the
short-distance mismatch between LGT and QCD, and tune it away.
It is natural to calculate the short-distance behavior with perturbative
QCD, owing to asymptotic freedom, and automated multi-loop calculations
for improved actions are feasible~\cite{Trottier:2003lt}.
Non-perturbative methods for matching are also
available~\cite{Bochicchio:1985za,Jansen:1996ck},
although most of the results apply only when $m_Qa\ll1$.
Combinations, in which some (or most) of the renormalization is computed
non-perturbatively
\cite{Hashimoto:1999yp,El-Khadra:2001rv,Hashimoto:2001nb}
are possible too.
For example, the normalization factor $Z_V(m_Qa)$ for the vector current
is easy to compute non-perturbatively for all $m_Qa$.
Then $\rho_A=Z_A/Z_V$ is tadpole-improved, and explicit
examples~\cite{Harada:2001fi} exhibit very small one-loop corrections and
mild mass dependence.


With these ideas in mind, we can summarize quickly the most common
methods for lattice $B$ physics, in each case working across
Table~\ref{tbl:methods}.

The first method~\cite{Gavela:1987rx,Bernard:1987pr} dates back to the
1987 Seillac conference.
It starts with Wilson quarks, nowadays almost always with the improved
clover action~\cite{Sheikholeslami:1985ij}.
Cutoff effects are described through the original Symanzik \LEL\ with
$m_Qa\ll1$.
The advantage is that non-perturbative matching calculations derived for
light quarks~\cite{Jansen:1996ck} may be taken over.
The disadvantage is that the heavy-quark mass in the computer is
artificially small (to keep $m_Qa<1$), so matrix elements of heavy-light
hadrons are extrapolated in $1/m_Q$ from $m_Q\sim m_c$ up to~$m_b$.
This method does not work for heavy-heavy systems, because $m_Q$
dependence in these systems, from NRQCD, is not characterized by powers
of~$1/m_Q$.

There are several names for this method in the literature.
Adherents prefer ``the relativistic method'' or, simply,~``QCD.''
The first name overlooks the major discretization effect, which stems
from violations of relativistic invariance.
The second is not helpful: all methods start with lattice gauge theory,
and use an effective field theory to define the matching and to interpret
cutoff effects.
Especially after the extrapolation in $1/m_Q$, this technique is not 
closer than the others to continuum QCD.
In the rest of this paper, I~call this method 
``the extrapolation method.''

At Seillac, Eichten introduced the method now known as
lattice HQET~\cite{Eichten:1987xu}.
Indeed, this work spawned the continuum HQET.
It starts with a discretization of the static approximation.
Then $1/m_Q$ corrections are treated as insertions~\cite{Eichten:1990vp}.
The Symanzik \LEL\ is then nothing but the continuum 
HQET~\cite{Kurth:2000ki}.
Non-perturbative renormalization methods have been worked out in the 
static limit~\cite{Heitger:2003xg,Hashimoto:pe}
but are sketchy for $1/m_Q$ insertions.
The static approximation and, hence, lattice HQET do not apply to 
quarkonium.

Another technique, also introduced at Seillac, is called lattice 
NRQCD~\cite{Lepage:1987gg}.
It first derives a non-relativistic effective Lagrangian for heavy quarks
in the continuum, and then discretizes.
The deviation from QCD, both from the non-relativistic expansion and from
the discretization, can be described with continuum heavy-quark theory:
HQET for heavy-light or NRQCD for heavy-heavy systems.
For the latter several interactions are needed (for few-percent
accuracy), so the lattice action can be rather
complicated~\cite{Lepage:1992tx}.
For this reason most of the matching calculations are perturbative.
The same lattice action is used for heavy-light and heavy-heavy systems,
so the latter can be used to test its accuracy, before proceeding to $B$
and $D$~physics.

The Fermilab method~\cite{El-Khadra:1996mp} is a synthesis.
Like the extrapolation method, it starts with (improved) Wilson fermions.
It confronts heavy-quark discretization effects in two ways.
One is to extend the Symanzik effective field theory to the regime
$m_Qa\not\ll1$.
The other is to use (continuum) HQET and NRQCD to describe both short 
distances, $a$ and $1/m_Q$.
This is possible because Wilson fermions have the right heavy-quark 
symmetries.
Renormalization and matching is usually partly non-perturbative and
partly perturbative.
Like lattice NRQCD, the Fermilab method can be used for heavy-light
and for heavy-heavy systems.

Heavy-light hadrons also contain light valence quarks.
In the past, Wilson quarks were usually used, but Wingate {\it et al.}\
have shown the advantages of naive quarks~\cite{Wingate:2002fh}.
Several difficulties associated with doublers are absent
because the naive quark is bound to a heavy quark.
Naive quark propagators are obtained from staggered quark propagators
by undoing the spin diagonalization.
With staggered quarks one can reach much smaller light quark masses
($m_q\sim 0.2m_s$) than with Wilson quarks ($m_q\sim0.5m_s$)
\cite{Gottlieb:2003lt,Jansen:2003lt}.

The main issue with the light quarks is the chiral extrapolation.
Heavy-meson chiral perturbation theory describes the dependence of
heavy-light masses and matrix elements on the light-quark
mass~\cite{Burdman:gh}.
It can be extended to partially quenched
simulations~\cite{Savage:2001jw}. 
For quarkonium the chiral extrapolation is less well developed.
The virtual process  most sensitive to the light-quark mass is
dissociation, which could have a strong effect on states just below 
threshold, such as $\psi'\to D\bar{D}\to\psi'$.
This process is completely omitted in the quenched approximation.

To close this section, let me list the (perceived) problems with the four
main treatments of heavy quarks.
The extrapolation method must balance the contradictory requirements of 
keeping both $\LQCD/m_Q$ and $m_Qa$ small.
In practice it is not clear that both sources of uncertainty, and their
interplay, are controlled.
In lattice HQET or lattice NRQCD, power-law divergences arise from 
the short-distance behavior of higher-dimension interactions and remain 
at some level when perturbative matching is used.
There are also the usual uncertainties from truncating perturbation
theory.
The Fermilab method, being based on Wilson fermions, does not have 
power-law divergences, but the perturbative part of the matching is 
said to suffer from a remnant called ``renormalon shadows.''
Another problem is that the $a$ dependence, though smooth, is not easily
described in the regime $m_Qa\sim1$.
These problems are assessed at the end of the next section, 
after discussing the theory of cutoff effects in detail.

\section{CUTOFF EFFECTS}
\label{sec:cutoff}

In this section I would like to discuss heavy-quark discretization
effects in more detail.
To do so, I shall use effective field theories to separate short- and
long-distance dynamics.
Discretization effects are at short distances, and here the methods
differ.
One can then assess the uncertainties from the short-distance behavior,
without precisely knowing the long-distance effects.

Although effective field theories appear in many contexts, 
it is perhaps worth recalling some of the basics.
At energies~$\Lambda$ below some scale~$\mu$, particles with $E>\mu$ have
small effects, suppressed by $(\Lambda/E)^n$.
The Coleman-Norton theorem~\cite{Coleman:1965le} shows that analytic
properties of Green functions are impervious to off-shell particles.
This is because singularities appear only where particles go on shell.
Suppose a heavy quark zig-zags into an anti-quark (emitting a hard
quantum), and then emits a soft quantum before turning back into a 
quark (absorbing a hard quantum).
The anti-quark is far off shell.
The analytic structure is retained even if the anti-quark and hard 
propagators are reduced to a point.
The resulting reduced diagram already looks like an 
effective-field-theory diagram.
New fields can be introduced to describe the remaining particles
and, hence, their non-analyticity.
To compensate for the omitted, off-shell particles, vertices
require generalized couplings.
This framework suffices, because field theory gives a complete
description respecting analyticity, unitarity, and the underlying
symmetries~\cite{Weinberg:1978kz}.

Asymptotic freedom provides another way to establish effective field
theories for underlying gauge theories, namely to all orders in the gauge
coupling.
In this way we have a rigorous proof (for Wilson quarks) of Symanzik's
theory of cutoff effects~\cite{Reisz:1988kk}, and a (less rigorous) proof
of the static field theory on which HQET is based~\cite{Grinstein:1990mj}.
Given the general arguments~\cite{Coleman:1965le,Weinberg:1978kz}
sketched above, it is hard to see where these would fail
non-perturbatively.
In particular, it seems implausible that confinement would be different
with two-component heavy-quark fields instead of four-component Dirac
fields.

With heavy quarks $m_Q\gg\LQCD$, so
zig-zags and pair production are suppressed in bound states.
Two-component fields $h^{(+)}_\vlabel$ ($h^{(-)}_\vlabel$) suffice to 
describe quarks (anti-quarks).
Here $\vlabel$ is a label, denoting a four-velocity close to that of 
the hadron containing the heavy quark(s).
In systems with one heavy quark, the heavy quark's velocity deviates 
from $\vlabel$ by an amount of order $\LQCD/m_Q$.
The field theory describing this physics with $h^{(\pm)}_\vlabel$ fields
is HQET.
If there are two heavy quarks, one has a binary system.
The two heavy quarks rotate slowly about their center of mass,
generating long-distance scales of order~$m_Q\nrvel^n$, 
where~$\nrvel\sim\alpha_s(m_Q)$ denotes the relative three-velocity.
In the $\bar{b}b$ ($\bar{c}c$) system $\nrvel^2\approx0.1$~($0.3$)
\cite{Lepage:1992tx}.
Now the field theory with $h^{(\pm)}_\vlabel$ fields is NRQCD.

HQET and NRQCD are based on an equivalence between QCD and an effective
Lagrangian.
One may write
\begin{equation}
	\mathcal{L}_{\mathrm{QCD}} \doteq 
		\mathcal{L}_{\mathrm{light}} + \mathcal{L}_{\mathrm{HQ}},
\end{equation}
where the symbol $\doteq$ can be read ``has the same matrix elements as.''
We cannot use $=$ (``equals''), because $\mathcal{L}_{\mathrm{QCD}}$
has Dirac fields for heavy quarks and $\mathcal{L}_{\mathrm{HQ}}$ has
heavy-quark fields $h_{\vlabel}^{(\pm)}$.
The effective Lagrangian
\begin{equation}
	\mathcal{L}_{\mathrm{HQ}} = 
		\sum_i \mathcal{C}_i(m_Q, m_Q/\mu) \mathcal{O}_i(\mu),
	\label{eq:LHQ}
\end{equation}
where the $\mathcal{C}_i$ are couplings, also known as short-distance
coefficients.
The operators $\mathcal{O}_i$ describe long-distance dynamics and, thus, 
bring in soft scales $\LQCD$, $m_Q\nrvel$, \emph{etc}.
The renormalization scale of the effective theory is $\mu$, chosen so 
that $(\LQCD,\,m_Q\nrvel)<\mu\lesssim m_Q$.
The difference between HQET and NRQCD
reflects the physical differences between heavy-light and heavy-heavy
systems, classifying the terms in Eq.~(\ref{eq:LHQ}) according to
powers of the respective small parameters,
$\LQCD/m_Q$ and~$(m_Q\nrvel)/m_Q=\nrvel$.

It is easiest to see the difference between the two by looking at the
leading and next-to-leading terms.
Let us choose the frame $\vlabel=(1,\bm{0})$ and, for brevity, suppress
the velocity label.
For HQET one counts powers of $\LQCD/m_Q$,
\begin{eqnarray}
	\mathcal{L}^{(0)}_{\mathrm{HQ}} & = & - \bar{h}(m_1 + D_4) h , 
	\label{eq:HQET:L0} \\
	\mathcal{L}^{(1)}_{\mathrm{HQ}} & = &  \displaystyle
		\frac{\bar{h}\bm{D}^2 h}{2m_2} + 
		z_\mathcal{B} \frac{\bar{h} i\bm{\Sigma}\cdot\bm{B} h}{2m_2} , \\
	\mathcal{L}^{(2)}_{\mathrm{HQ}}	& = &  \displaystyle
		z_{\mathrm{D}} \frac{\bar{h} \bm{D}\cdot\bm{E}h}{8m_2^2} +
		z_{\mathrm{s.o.}}
		\frac{\bar{h}i\bm{\Sigma}\cdot[\bm{D}\times\bm{E}]h}{8m_2^2} , 
	\label{eq:HQET:L2}
\end{eqnarray}
and in general $\mathcal{L}^{(s)}_{\mathrm{HQ}}$ consists of all 
interactions of dimension $s+4$.
The $z_i$ are dimensionless versions of the $\mathcal{C}_i$;
they depend on $m_Q/\mu$ with anomalous dimensions.
Thus, HQET has both power-law and logarithmic $m_Q$ dependence.

For NRQCD, on the other hand, one counts powers of~$\nrvel$.
As explained in Ref.~\cite{Lepage:1992tx}, $\bm{D}\sim m_Q\nrvel$,
$\bm{B}\sim m_Q^2\nrvel^4$, $\bm{E}\sim m_Q^2\nrvel^3$.
Then
\begin{eqnarray}
	\mathcal{L}^{(0)}_{\mathrm{HQ}} & = &  - \bar{h}(m_1 + D_4) h + 
		\displaystyle \frac{\bar{h}\bm{D}^2 h}{2m_2} , 
	\label{eq:NRQCD:L0} \\
	\mathcal{L}^{(2)}_{\mathrm{HQ}} & = & \displaystyle
		z_\mathcal{B} \frac{\bar{h} i\bm{\Sigma}\cdot\bm{B} h}{2m_2} +
		z_{\mathrm{D}} \frac{\bar{h} \bm{D}\cdot\bm{E}h}{8m_2^2} +
	\label{eq:NRQCD:L2} \\
	&  & \displaystyle 
		z_{\mathrm{s.o.}}
		\frac{\bar{h}i\bm{\Sigma}\cdot[\bm{D}\times\bm{E}]h}{8m_2^2} - 
		z_{\rm rel} \frac{\bar{h}(\bm{D}^2)^2h}{8m_2^3} .
		\nonumber 
\end{eqnarray}
In general $\mathcal{L}^{(s)}_{\mathrm{HQ}}$ consists of all terms that
scale like~$\nrvel^{s+2}$.
Eqs.~(\ref{eq:NRQCD:L0}) and~(\ref{eq:NRQCD:L2}) contain the same terms
as Eqs.~(\ref{eq:HQET:L0})--(\ref{eq:HQET:L2}), \emph{etc.}, but rearranged.
For example, the kinetic energy is an essential effect in NRQCD, 
but a non-leading correction in HQET.

The rest mass~$m_1$ and kinetic mass~$m_2$ should \linebreak be thought of as 
short-distance coefficients.
When describing QCD with ${\cal L}_{\mathrm{HQ}}$, $m_1=m_2$.
If the operators are renormalized in the $\overline{\rm MS}$ scheme, 
this mass is (in perturbation theory) the pole mass.
When we turn to LGT below, $m_1\neq m_2$.

I believe that the key to reliable calculations 
lies in firmly incorporating these heavy-quark ideas into LGT.
After all, almost nothing is known about heavy quarks (in bound states)
without these and other scale-separation techniques.
In particular, the machinery outlined here can be applied to any
underlying field theory with heavy-quark symmetry, including LGT with
Wilson, static, or non-relativistic quarks.
Many readers will agree with such a strategy, but others seem to
believe that effective theories should be avoided.
If so, they overlook the fact that the alternative theory of cutoff
effects---Symanzik's---is (!)\ an effective field theory.

Let us turn, then, to the Symanzik theory, with an eye to nuances 
that arise for heavy (Wilson) quarks.
To describe cutoff effects, Symanzik~\cite{Symanzik:1979ph}
introduces a local effective Lagrangian~(\LEL).
We can write
$\mathcal{L}_{\mathrm{LGT}}\doteq\mathcal{L}_{\mathrm{Sym}}$, and
\begin{eqnarray}
	\mathcal{L}_{\mathrm{Sym}} & = & 
		\frac{1}{2g^2}\tr[F_{\mu\nu}F^{\mu\nu}] - 
		\sum_f \bar{q}_f(\Dirac + m_f)q_f 
		\label{eq:Sym} \\[-3pt]
		& + & \sum_i 
		K_i(a, g^2, \{m_fa\}; \{c_j\}; \mu a) O_i(\mu).
		\nonumber
\end{eqnarray}
The first line gives continuum QCD, with renormalized coupling~$g^2$ and
quark masses $m_f$.
The sum in Eq.~(\ref{eq:Sym}) accounts for cutoff effects.
Quarks, even heavy ones, are described by fields $q_f$
satisfying the Dirac equation.
The renormalization scale $\mu$ separates the short distance~$a$
from long distances, principally~$1/\LQCD$.
For a light quark, $1/m_f$ is another long distance, so it is sensible 
and accurate to expand the coefficients $K_i$ in $(m_fa)^n$.
In this way one obtains an expansion in~$a$ that is, however, a
consequence of separating scales when the only short distance is~$a$.

For a heavy quark~$Q$, on the other hand, $1/m_Q$ is (compared to $1/\LQCD$)
a short distance.
Hence, it does not make sense to expand the $K_i$ in $m_Qa$.
The expansion is also inaccurate when $m_Qa\not\ll1$.
Instead of expanding, one should rearrange the terms in Eq.~(\ref{eq:Sym})
to collect the largest terms from the sum.
They are of the form $\bar{Q}X(\gamma_4D_4)^nQ+{\rm h.c.}$, which can be
eliminated with the equations of motion in favor of explicit mass 
dependence and soft
terms~\cite{El-Khadra:1996mp,Aoki:2001ra,Kronfeld:2002pi}.
One finds
\begin{eqnarray}
	\mathcal{L}_{\rm Sym} & = & \mathcal{L}_{F,q} -
		\bar{Q}\left(m_1 + \gamma_4D_4 + 
		\sqrt{\frac{m_1}{m_2}}\bm{\gamma}\!\cdot\!\bm{D} \right)Q
	\nonumber \\
		& + & \textrm{small corrections},
	\label{eq:mySym}
\end{eqnarray}
The incorrect normalization of the spatial kinetic energy captures the 
breaking of relativistic invariance when $m_Qa\not\ll1$.
This modified \LEL\ is on the same footing as Eq.~(\ref{eq:Sym}).
Thus, one sees that for $m_Qa\not\ll1$ it is neither lattice gauge theory
nor the Symanzik \LEL\ that breaks down.
Instead, the split ``QCD + small corrections'' suggested in 
Eq.~(\ref{eq:Sym}) is lost in Eq.~(\ref{eq:mySym}).

The apparent obstacle can be overcome by using HQET and NRQCD to
separate the short distances $(a,\,1/m_Q)$ from $(\LQCD,\,m_Q\nrvel^n)$.
The logic and structure for LGT is the same as for
QCD \cite{Kronfeld:2000ck}.
With two short distances, the Wilson coefficients depend on the
dimensionless ratio $m_Qa$:
\begin{eqnarray}
	\mathcal{C}_i(m_Q;\mu/m_Q) \to 
		\mathcal{C}_i(m_Q, m_Qa; \{c_j\};\mu/m_Q) .
\end{eqnarray}
As shown, they also depend on improvement couplings~$c_j$ in the lattice
action, as do the coefficients $K_i$ in Eq.~(\ref{eq:Sym}).

The rest mass $m_1$ and kinetic mass $m_2$ in Eq.~(\ref{eq:mySym}) are
the same as Eqs.~(\ref{eq:HQET:L0})--(\ref{eq:NRQCD:L0}).
But in HQET and NRQCD $m_1$ multiplies the conserved number operator 
$\bar{h}h$: it contributes additively to the mass spectrum but not at 
all to matrix elements.
So, although it is possible to introduce a new parameter to tune 
$m_1=m_2$~\cite{El-Khadra:1996mp}, 
it is also unnecessary (assuming $m_Q\gg\LQCD$).
In this way, the heavy-quark theories make sense out of the short-distance
behavior of Wilson quarks, where the Symanzik theory does not.

With HQET and NRQCD descriptions of static, non-relativistic, and Wilson 
quarks, we are in a position to make comparisons.
Heavy-quark expansions for QCD and LGT can be developed
side-by-side~\cite{Kronfeld:1995nu}, showing
explicitly~\cite{Harada:2001fi} that heavy-quark discretization effects
are lumped into 
\begin{equation}
	\delta{\cal C}_i = 
		{\cal C}^{\rm LGT}_i(\{c_j\}) - {\cal C}^{\rm QCD}_i .
	\label{eq:deltaC}
\end{equation}
Solving the equations $\delta{\cal C}_i=0$ for the
couplings~$c_j$ yields on-shell improvement conditions.
For currents there is an analogous description leading to similar 
mismatches and to normalization factors
\begin{equation}
	Z_J=C^{\rm QCD}_J/C^{\rm LGT}_J ,
	\label{eq:Z}
\end{equation}
where $C_J^{\cdots}$ is the matching factor from the underlying theory
to the effective theory.
The resulting expressions for the $c_j$ and $Z_J$ no longer depend on the
renormalization scheme of ${\cal L}_{\rm HQ}$:
they match LGT directly to~QCD.

To turn these results into semi-quantitative estimates, one can look at
the mismatches in the coefficients.
Let us focus on the action for illustration.
In the extrapolation method the quark mass is identified with the rest
mass, $m_Q=m_1$, so the kinetic energy introduces an error
\begin{equation}
	\delta\mathcal{C}_2^\mathrm{(extrapolation)} \langle O_2\rangle = 
		\left|\frac{p}{2m_2} - 
		\frac{p}{2m_1}\right|_{m_Q=m_1}\!,
	\label{eq:extrapError}
\end{equation}
where $p$ is a soft scale.
The chromomagnetic mismatch 
$\delta\mathcal{C}_\mathcal{B}\langle O_\mathcal{B}\rangle$ is similar.
In lattice NRQCD the quark mass is adjusted non-perturbatively to the 
kinetic energy: $m_Q=m_2$.
The first error is in spin-dependent effects.
With $l$-loop matching
\begin{equation}
	\delta\mathcal{C}_\mathcal{B}^\mathrm{(NRQCD)} 
		\langle O_\mathcal{B}\rangle \sim 
		\alpha_s^{l+1}\left(1+\frac{1}{4m_Q^2a^2}\right) \frac{p}{2m_Q}
	\label{eq:NRQCDError}
\end{equation}
has the right asymptotics in~$m_Qa$.
The power-law divergence comes from the short-distance part of 
higher-dimension spin-dependent interactions.
The Fermilab method chooses $m_Q=m_2$ and, so, eliminates the 
error~(\ref{eq:extrapError}).
The leading error is 
\begin{equation}
	\delta\mathcal{C}_\mathcal{B}^\mathrm{(Fermilab)} 
		\langle O_\mathcal{B}\rangle \sim 
		\left.\frac{\alpha_s^{l+1}pa}{2(1+m_0a)}\right|_{m_Q=m_2}\!,
	\label{eq:FermiError}
\end{equation}
again with $l$-loop matching of the chromomagnetic energy.
The existence of a continuum limit controls the $a\to 0$ limit.

\begin{figure}[b!p]
	\vspace{-24pt}
	\includegraphics[width=75mm]{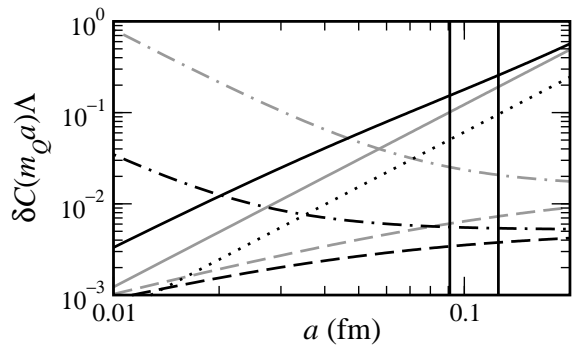}
	\vspace{-36pt}
	\caption{Comparison of errors.
	Black: $Q=b$;
	gray: $Q=c$.
	Solid: extrapolation method;
	dashed-dotted: lattice NRQCD;
	dashed: Fermilab method;
	dotted: gluons and light quarks.}
	\label{fig:errors}
\end{figure}

The error estimates of Eqs.~(\ref{eq:extrapError})--(\ref{eq:FermiError})
are plotted in Fig.~\ref{fig:errors}, taking $p=700$~MeV $(\sim\LQCD)$,
$m_c=1400$~MeV, $m_b=4200$~MeV, $\alpha_s=0.25$, and $l=1$.
The vertical lines indicate the lattice spacings of the unquenched MILC 
ensembles~\cite{Gottlieb:2003lt,Bernard:2001av}.
The results substantiate some of the perceived problems.
For both charmed quarks (gray curves) and for bottom quarks (black
curves), the extrapolation method has largest discretization effects,
10--20\% at feasible lattice spacing.
Lattice NRQCD (with perturbative matching) has an increasing error as
$a\to0$, as noted in the first paper on NRQCD~\cite{Lepage:1987gg}.
The Fermilab method's heavy-quark discretization effects are not a pure
power of~$a$ at feasible lattice spacings, where $m_Qa\sim1$.

Calculations of matrix elements have similar errors from matching and 
further errors from the normalization factors.
With lattice NRQCD and the Fermilab method the latter are, till now,
of order $\alpha_s^2$ from one-loop matching.
Two-loop matching is underway~\cite{Trottier:2003lt}.
Matrix elements from the extrapolation method typically have leading 
normalization errors of order $\alpha_sm_Q^2a^2$.
Some of the ``improvement'' contributions, taken over from light 
quarks, diverge in the heavy-quark limit.
They are worrisome and should be avoided.

Fig.~\ref{fig:errors} reveals two other interesting features.
First, the extrapolation method has smaller discretization effects than
lattice NRQCD only at unfeasibly small lattice spacings.
Second, discretization errors from the light quarks, here taken to be 
$\frac{1}{2}(pa)^2$, seem to dominate in lattice NRQCD and the 
Fermilab method.
That suggests that one could carry out a continuum extrapolation to get
rid of these effects, tolerating a percent-level bias from the heavy
quarks.

One can take care of the discretization effects of the extrapolation 
method by taking the continuum limit first~\cite{Rolf:2003mn}, but it is
dangerous.
On the lattices available today (especially in unquenched simulations),
one is driven to artificially small values of the heavy-quark mass.
At some point, however, the $1/m_Q$ expansion breaks down.
The threshold effects that allow pair production and zig-zags are smeared
out by the momentum distribution of gluons inside the hadron, so,
unfortunately, the breakdown will not be obvious.

The above discussion skips over lattice HQET.
With perturbative matching, the kinetic, as well as the chromomagnetic, error
is like~(\ref{eq:NRQCDError}).
For this and signal-to-noise reasons, this method has lately fallen from 
favor.
The latter problem now seems to be solved~\cite{DellaMorte:2003mn}, and 
there is progress on non-perturbative matching~\cite{Heitger:2003xg}.
The way I have set up HQET makes it clear, however, that 
non-perturbative matching, say in a volume of finite size~$L$, should work 
with minor modifications for static, non-relativistic, and Wilson quarks.
Non-perturbative matching in finite volume will introduce
uncertainties of order $\LQCD/m_Q^2L$.
To control these $L$ cannot be too small.
\vspace{-2pt}
\section{RENORMALON SHADOWS}
\label{sec:renshad}

We now turn to renormalon shadows~\cite{Bernard:2000ki}, which are 
perceived to be a problem in the Fermilab method.

Renormalons are power-law ambiguities that arise in large orders of
perturbation theory, when mass-independent renormalization schemes are
used.
Several years ago, Martinelli and Sachrajda considered the problem of 
power corrections~\cite{Martinelli:1996pk}.
They concluded that a sluggish cancellation of renormalon effects could 
make it difficult to take power corrections into account.
Frankly, I am suspicious of sweeping conclusions that hinge on
renormalons, because in other renormalization schemes they do not
arise~\cite{Bodwin:1998mn}.
Let us see, therefore, whether it is possible to arrive at similar 
conclusions without reference to renormalons.

Suppose one can measure experimentally (or compute non-perturbatively)
$\mathcal{P}$ and $\mathcal{Q}$.
Both are described in an effective field theory including a power
correction:
\begin{eqnarray}
	\mathcal{P}(Q) & = & C_{\cal P}(Q/\mu) \langle O_1\rangle + 
		B_{\cal P}(Q/\mu) \langle O_2\rangle/Q , \label{eq:P} \\
	\mathcal{Q}(Q) & = & C_{\cal Q}(Q/\mu) \langle O_1\rangle + 
		B_{\cal Q}(Q/\mu) \langle O_2\rangle/Q , \label{eq:Q}
\end{eqnarray}
where the $C$s and $B$s are short-distance coefficients.
$Q$~is a hard physical scale, and $\mu$ is the separation scale.
By assumption, the same effective-theory matrix elements 
$\langle O_1\rangle$ and $\langle O_2\rangle$ appear in 
$\mathcal{P}$ and $\mathcal{Q}$.
Now one would like to predict
\begin{equation}
	\mathcal{R}(Q) = C_{\cal R}(Q/\mu) \langle O_1\rangle + 
		B_{\cal R}(Q/\mu) \langle O_2\rangle/Q ,
\end{equation}
given $\mathcal{P}$, $\mathcal{Q}$, and approximate short-distance
coefficients.
What is the uncertainty in $\mathcal{R}$?

One could solve Eqs.~(\ref{eq:P}) and (\ref{eq:Q}) for the 
$\langle O_i\rangle$, but renormalons could enter through the 
renormalization scheme for the effective field theory.
Instead, just use linear algebra to obtain: 
\begin{equation}
	\mathcal{R} \!=\! \frac{C_{\cal R}}{2}\!\!
	\left[\frac{\cal P}{C_{\cal P}}\!+\!\frac{\cal Q}{C_{\cal Q}} \right] +
	\frac{\bar{B}_{\cal R}[\mathcal{P}/C_{\cal P} - \mathcal{Q}/C_{\cal Q}]}%
	{B_{\cal P}/C_{\cal P} - B_{\cal Q}/C_{\cal Q}},
	\label{eq:eft-w/o-eft}
\end{equation}
where $\bar{B}_{\cal R}$ is a combination of the $B$s and $C$s.
The second term is the power correction: the combination
$\mathcal{P}/C_{\cal P} - \mathcal{Q}/C_{\cal Q}$ 
is formally of order~$\LQCD/Q$.

To estimate the uncertainty in~$\mathcal{R}$, let us assume that the~$C$s
have been calculated through $l$ loops and the $B$s through $k$ loops.
Then the customary assumption is that the uncertainty in the (leading-twist)
first bracket is $O(\alpha_s^{l+1})$ and in the power correction
$O(\alpha_s^{k+1}\LQCD/Q)$.
Moreover, it seems worthwhile to include the power correction as soon as
$\LQCD/Q\gtrsim\alpha_s^{l+1}$.

But the leading-twist parts of
$\mathcal{P}/C_{\cal P}-\mathcal{Q}/C_{\cal Q}$ do not cancel perfectly
when $C_{\cal P}$ and $C_{\cal Q}$ are calculated perturbatively.
There is a shadow~\cite{Bernard:2000ki} of order $\alpha_s^{l+1}$,
commensurate with the uncertainty in the leading-twist term.
Unless $\alpha_s^{l+1}\ll\LQCD/Q$, the shadow obscures the desired
power correction, which is especially likely if $\mathcal{P}$ and
$\mathcal{Q}$ are dissimilar~\cite{Martinelli:1996pk}.
Then the second term in Eq.~(\ref{eq:eft-w/o-eft}) would not improve 
the accuracy of~$\mathcal{R}$.

To see if this reasoning applies to any of the heavy-quark methods, one
must check if the same kind of cancellation is needed.
In lattice HQET, where $1/m_Q$ corrections are insertions, the foregoing
analysis goes through without change, so the shadow is an issue.
The Fermilab method applies HQET in a different way, so the conclusions 
can be (and are) different.
The HQET description gives formulas of the 
form~\cite{Kronfeld:2000ck,Harada:2001fi}
\begin{eqnarray}
	\Phi^{\rm LGT} & = & 
		C^{\rm LGT} \left[\Phi_\infty + B^{\rm LGT}\Phi'_\infty/m_Q \right] ,\\
	\Phi^{\rm QCD} & = & 
		C^{\rm QCD} \left[\Phi_\infty + B^{\rm QCD}\Phi'_\infty/m_Q \right] .
\end{eqnarray}
We want to know the uncertainty, when $Z=C^{\rm QCD}/C^{\rm LGT}$ is
computed through $l$~loops, and $\delta B=B^{\rm LGT}-B^{\rm QCD}$ is
matched through $k$~loops.
One finds the relative error
\begin{equation}
	1 - \frac{Z^{(l)}\Phi^{\rm LGT}}{\Phi^{\rm QCD}} = 
		\frac{\delta Z^{(l+1)}}{Z} - 
			\frac{Z^{(l)}\delta B^{(k+1)}\Phi'_\infty}{Z\Phi_\infty m_Q} ,
\end{equation}
where $Z^{(l)}$ is the $l$-loop approximation to~$Z$, and
$\delta Z^{(l+1)}=Z-Z^{(l)}$.
The first term has truncation error $\alpha_s^{l+1}$, 
the second $\alpha_s^{k+1}\LQCD/m_Q$.
There is no shadow contribution, because the power correction is not
obtained by explicit subtraction.
Instead, it is present all along, and LGT is adjusted to hit
its target,~QCD.

One can always worry that the first uncalculated coefficient in
$Z^{(l+1)}$ or $\delta B^{(l+1)}$ is large.
But those are the errors discussed in Sec.~\ref{sec:cutoff}.
They have nothing to do with renormalons or shadows.

\section{A NEW TECHNIQUE}
\label{sec:new}

Among this year's several methodological developments,
I would like to discuss a new finite-volume 
technique~\cite{Guagnelli:2002jd,deDivitiis:2003iy}.
The idea is to calculate a heavy-light observable~$\Phi$ in a sequence of 
finite volumes of size $L_0$, $2L_0$, \ldots.
Then
\begin{equation}
	\Phi(\infty) = \Phi(L_0)\sigma(L_0)\sigma(2L_0) \cdots,
	\label{eq:TV}
\end{equation}
where
$\sigma(L) = \Phi(2L)/\Phi(L)$.
To see how it works, let us follow the published 
work~\cite{Guagnelli:2002jd,deDivitiis:2003iy}.
With $L_0=0.4$~fm the box is small enough so that $m_ba\ll1$ is 
feasible (though not cheap!).
Thus, $\Phi(L_0)$ can be obtained in the continuum limit.
One obtains $\sigma(2^{j-1}L_0)$ by taking the continuum limit first, 
and then extrapolating in $1/m_Q$ from $2^j/m_b$ to $1/m_b$.
``Infinite'' volume is reached after two steps, where $L=4L_0=1.6$~fm.

Systematic uncertainties stem from the $1/m_Q$ extrapolations.
Let us focus on meson masses, which, in HQET, are written
$M(L)=m+\bar{\Lambda}(L)$, where $m$ is the quark mass.
Renormalization-scheme dependence cancels between $m$ and $\bar{\Lambda}$.
$\bar{\Lambda}$ encodes the long-distance physics, so it depends on the 
box size~$L$, but $m$ does not.
Thus,
\begin{equation}
	\sigma(L) = 1 + 
		\left[\bar{\Lambda}(2L)-\bar{\Lambda}(L)\right]/m.
\end{equation}
Because of the extrapolation, one really has
\begin{equation}
	\sigma(2^{j-1}L_0) = 1 + \frac{2^j}{\mbox{``$2^j$''}}
	\frac{\bar{\Lambda}(2^jL_0)-\bar{\Lambda}(2^{j-1}L_0)}{m} ,
\end{equation}
where ``$2^j$'' expresses the uncertainty arising from the
extrapolation.
Because $\sigma$ is computed in the continuum limit, these uncertainties
arise from higher orders in $1/m$ (for $j=1$) and very possibly from a
breakdown of the heavy-quark expansion (for $j=2$).
When extrapolation is most dangerous (larger $j$), the difference
$\bar{\Lambda}(2^jL_0)-\bar{\Lambda}(2^{j-1}L_0)$ nearly vanishes,
\pagebreak
according to the usual asymptotic $L$ dependence of hadron masses.

The heavy-light decay constants can be analyzed similarly, but HQET
anomalous dimensions make the $1/m$ extrapolation less clean.
The cancellation mechanism still works, however.

For bottomonium the extrapolations do not work as above.
NRQCD says $M(L)=2m+B(L)$, where the binding energy 
$B\sim m\nrvel^2$, with relative velocity $\nrvel\sim\alpha_s(m)$.
Extrapolations linear or quadratic in $1/m$ are, thus, not motivated.

\section{SPECTRUM AND CKM RESULTS}
\label{sec:results}

This section surveys physics results, although less thoroughly than in
the past two years~\cite{Ryan:2001ej,Yamada:2002wh}.
It now makes sense to focus on unquenched calculations, which are just 
beginning.
Here I focus on the quarkonium spectrum, the new $D_s$ states, 
semi-leptonic decays, and $B$-$\bar{B}$ mixing.

\subsection{Quarkonium}

\begin{figure}[b!p]
	\vspace{-12pt}
	\includegraphics[width=75mm]{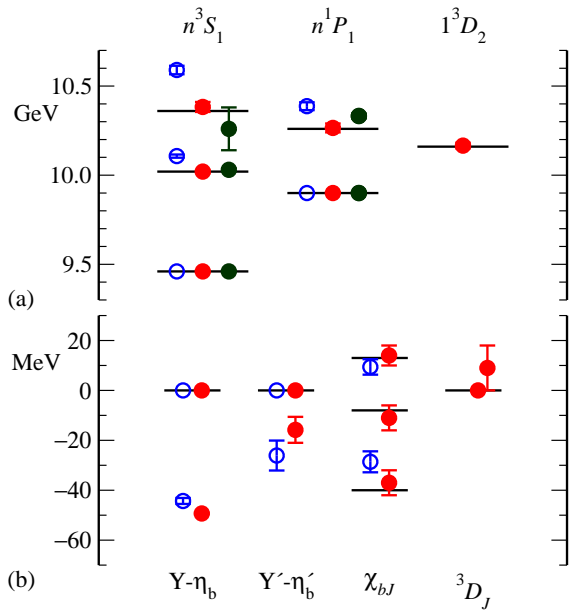}
	\vspace{-36pt}
	\caption[fig:ups]{Bottomonium spectrum~\cite{Davies:2003pc}.
	Open (closed) symbols show quenched ($2+1$) results. 
	(a) Gross structure with lattice spacings
	$a=\frac{1}{8}$~fm (middle) and $a=\frac{1}{11}$~fm (right).
	(b) Spin-dependent splittings with $a=\frac{1}{8}$~fm.}
	\label{fig:ups}
\end{figure}

Fig.~\ref{fig:ups} shows the gross and fine structure of 
bottomonium~\cite{Davies:2003pc}, 
comparing quenched results to those on the MILC
ensembles~\cite{Gottlieb:2003lt,Bernard:2001av} with $2+1$ flavors.
The lattice spacing is fixed by the spin-averaged $1P$-$1S$ splitting,
and the overall mass by the $\Upsilon(1S)$ level.
In both parts of Fig.~\ref{fig:ups} one sees much better agreement with 
experiment for $2+1$ sea quarks.
Note particularly the improvement in the excited levels $2S$, $3S$, 
and $2P$, and the improvement in the $\chi_{bJ}$ splittings.

We have similar results charmonium too.
In this case, we can compare the hyperfine splitting 
$m_{J/\psi}-m_{\eta_c}$ to experiment,
finding an $18\pm2\%$ deviation~\cite{Simone:2003tp}.
This is better than in the quenched approximation, but still 
unsatisfactory.
Note, however, that the heavy-quark chromomagnetic interaction is 
matched only at the tree level.
One-loop matching~\cite{Nobes:2003nc,Aoki:2003ka} should fix
the problem.

The unquenched quarkonium spectrum allows us to fix the scales $r_0$ 
and $r_1$ from first principles.
Calculating the heavy-quark potential and fixing lattice units from 
the $\Upsilon$ spectrum, the MILC and HPQCD collaborations 
find~\cite{MILC:2003pc}
\begin{equation}
	r_0 = 0.46 (1)~\mathrm{fm}, \qquad r_1 = 0.32 (1)~\mathrm{fm},
	\label{eq:r0}
\end{equation}
at both lattice spacings, $\frac{1}{8}$ and $\frac{1}{11}$~fm.
Eq.~(\ref{eq:r0}) is the first QCD determination of $r_0$; 
the value $r_0=0.5$~fm \cite{Sommer:1993ce} comes from potential models.

\subsection{\boldmath $D_s(0^+)$ and $D^*_s(1^+)$}
\label{subsec:Ds}

Now let us turn to the new positive-parity $D_s$ 
states~\cite{Aubert:2003fg,Besson:2003cp,Abe:2003jk}.
The experiments find~\cite{Besson:2003cp}
\begin{eqnarray}
	m_{D_s}(0^+)  -m_{D_s}(0^-)   & = & 350(1)~\textrm{MeV}, \\
	m_{D^*_s}(1^+)-m_{D^*_s}(1^-) & = & 351(2)~\textrm{MeV}.
\end{eqnarray}
These observations have inspired several lattice calculations~%
\cite{Bali:2003jv,Dougall:2003hv,Mackenzie:2003hq}.
The UKQCD Collaboration includes a simulation with
$n_f=2$, $m_q^{\rm sea}\approx m_s$, at fixed lattice spacing,
$a\approx 1/10$~fm \cite{Dougall:2003hv}.
Ref.~\cite{Dougall:2003hv} converts from lattice units to physical
units with $r_0=0.55$~fm.
With $r_0=0.46$~fm, from Eq.~(\ref{eq:r0}), their unquenched results
become~\cite{Maynard:2003hv}
\begin{eqnarray}
	m_{D_s}(0^+)-m_{D_s}(0^-) & = & 494(28)~\textrm{MeV}, \\
	m_{D_s}(1^+)-m_{D_s}(1^-) & = & 445(28)~\textrm{MeV},
\end{eqnarray}
while $m_{D_s^*}(1^-)-m_{D_s}(0^-)=140(3)$~MeV 
agrees with experiment.
(Errors from non-zero~$a$ and imperfect unquenching are not reported here.)

These splittings are too high, but in this case that is what one should
expect.
In Nature the threshold $D^{(*)}K$ lies just above $D^{(*)}_s(J^+)$, 
so the virtual process $D_s(J^+)\to D^{(*)}(J^-)K\to D_s(J^+)$ pushes
$m_{D^{(*)}_s}(J^+)$ down.
UKQCD's sea-quark mass $m_q^{\rm sea}$ is slightly higher than the
physical strange quark mass, so the threshold in simulation is 
higher, diluting its importance.
If $m_q^{\rm sea}$ would be reduced (eventually to $m_d$), the threshold 
effect would reappear.
Some evidence for this mechanism comes from the MILC
ensembles~\cite{Gottlieb:2003lt,Bernard:2001av} with lighter sea quarks,
where we preliminarily do find lower splittings~\cite{Mackenzie:2003hq}.

\subsection{Semi-leptonic Decays}
\label{subsec:semi}

Semi-leptonic decays determine the top two rows of the CKM matrix,
because they are unlikely to be sensitive to non-Standard physics.
Lattice QCD calculates the hadronic form factors, $f_+(q^2)$,
$f_+(E_\pi)$, $\mathcal{F}(w)$, \emph{etc.}, from matrix elements
$\langle\pi|V^\mu|K\rangle$, $\langle\pi|V^\mu|B\rangle$,
$\langle D^*|J^\mu|B\rangle$, \emph{etc}.

To determine $|V_{cb}|$ through the exclusive decay $B\to D^*l\nu$, 
one needs the zero-recoil form factor $\mathcal{F}(1)$.
Exact heavy-quark spin and flavor symmetries would imply 
$\mathcal{F}(1)=1$~\cite{Isgur:1989vq}.
The (quenched) state of the art is~\cite{Hashimoto:2001nb}
\begin{equation}
	\mathcal{F}(1)\!=\!0.913^{+0.024}_{-0.017}\pm0.016{}^{+0.003}_{-0.014}
		{}^{+0.000}_{-0.016}{}^{+0.006}_{-0.014},
\end{equation}
where the error bars are from statistics, 
HQET matching, non-zero lattice-spacing, chiral extrapolation,
and quenching.
The total uncertainty of~4\% hinges on the HQET description of LGT,
so all uncertainties scale with $\mathcal{F}(1)-1$, not~$\mathcal{F}(1)$.

How will these uncertainties change in future unquenched calculations?
The HQET matching error bar $\pm0.016$ is reducible through higher-order 
improvement~\cite{Oktay:2003gk}.
Of the rest, the chiral error bar ${}^{+0.000}_{-0.016}$ is the most 
subtle.
The $D^*$ in the final state is unstable and, hence, not clearly 
gold-plated.
In Nature, however, $m_{D^*}-m_D$ is only slightly larger than~$m_\pi$, 
so the $D^*$ is narrow,
$\Gamma_{D^*}\approx100$~keV~\cite{Anastassov:2001cw}.
In any feasible simulation, the artificially high mass of 
the light valence quark puts the $D^*_q$ above threshold.
This effect has been worked out in 
$\chi$PT~\cite{Randall:qg,Savage:2001jw} and, in fact, 
leads to the ${}^{+0.000}_{-0.016}$.
With more experience and, hence, confidence in chiral extrapolations, 
one may hope to reduce (or at least symmetrize) this uncertainty.
This would then be a 1\% uncertainty on~$|V_{cb}|$.

Charmless semi-leptonic decays can be used to determine~$|V_{ub}|$.
The CLEO Collaboration~\cite{Athar:2003yg} finds that $B\to\pi l\nu$ is
favored experimentally over $B\to\rho l\nu$.
That is fortunate for lattice QCD, because the $\pi$ is stable and the 
$\rho$ is not.
Several unquenched calculations are in progress~\cite{Shigemitsu:2003xw}.
The final-state pion can reach high energy, $E_\pi\le2.6$~GeV, which is
another source of discretization errors.
Ideas like moving NRQCD~\cite{Foley:2002qv,Boyle:2003ui} promise to 
manage it.
As usual one must worry about the chiral extrapolation.
A nice LGT-oriented guide~\cite{Becirevic:2002sc} for low~$E_\pi$ comes
from (partially quenched) $\chi$PT.
One would like to treat energetic final-state pions as well,
which, as far as I know, is an open problem.

\subsection{\boldmath $f_B$ and $\bar{B}^0_q$-$B^0_q$ Mixing}

Neutral $\bar{B}^0_q$-$B^0_q$ mixing plays a central role in $B$ physics.
(Here $q$ labels the spectator $d$ or~$s$ quark).
The theoretical expression for the mixing frequency can be related 
to physics at the electroweak scale (and beyond), if one knows a QCD
quantity called $\eta_B{\cal M}_q$.
Here
\begin{equation}
	{\cal M}_q=\langle\bar{B}_q^0| [\bar{b}\gamma^\mu (1-\gamma^5)q]
		[\bar{b}\gamma_\mu (1-\gamma^5)q]|B_q^0\rangle,
\end{equation}
and $\eta_B$ is a short-distance coefficient.
$\mathcal{M}_q$ and $\eta_B$ depend on the scheme for integrating out
$W$, $Z$, and~$t$, but $\eta_B{\cal M}_q$ does not.
The renormalization-group invariant value of the short-distance factor
is $\hat{\eta}_B=0.55$.

For historical reasons one usually writes
\begin{eqnarray}
	B_{B_q} & = & 3{\cal M}_q/8m^2_{B_q}f^2_{B_q} , \\
	f_{B_q} & = & \langle0|\bar{b}\gamma_\mu\gamma_5 q|B_q^0\rangle/m_{B_q}.
\end{eqnarray}
The separation into the decay constant $f_{B_q}$ 
and bag factor $B_{B_q}$ turns out to be useful.

$\Delta m_d$ is precisely measured~\cite{Hagiwara:fs}, and $\Delta m_s$
will be soon~\cite{Anikeev:2001rk}.
It is tempting to test the CKM matrix with $\Delta m_s/\Delta m_d$.
Then lattice QCD should provide the ``SU(3)-breaking'' ratios
$\mathcal{M}_s/\mathcal{M}_d$ or
\begin{equation}
	\xi = \xi_f \xi_B, \hfill \xi_f = f_{B_s}/f_{B_d}, 
	\hfill \xi^2_B=B_{B_s}/B_{B_d}. \hfill
\end{equation}
For several years, the conventional wisdom held that $\xi$ (and $\xi_f$) 
would enjoy large cancellations in the uncertainties.
Several uncertainties do cancel in the ratios, but one of the 
largest---from the chiral extrapolation of $\xi_f$---does not.

A correct way to carry out the chiral extrapolation is, of course, 
to use~$\chi$PT.
In the case at hand, $\chi$PT provides the description
$f_{B_q}=F[1+\Delta f_q]$, $B_{B_q}=B[1+\Delta B_q]$, 
where $F$ and $B$ are constants and $\Delta f_q$ and $\Delta B_q$ are 
calculated from loop diagrams.
For the $B_d$ meson one finds~\cite{Booth:1994hx}
\begin{eqnarray}
	\Delta f_d & = & -G_+
		\left[\case{3}{4} I(m_\pi^2) + \case{1}{2} I(m_K^2) + 
		\case{1}{12} I(m_\eta^2)\right] \nonumber \\
		& & +\,m_K^2 f_1 + \case{1}{2}m_\pi^2 [f_1 + f_2], 
	\label{eq:unquenchedfBd} \\
	\Delta B_d & = & -G_-
		\left[\case{1}{2} I(m_\pi^2) + \case{1}{6} I(m_\eta^2)\right]
		 + m_K^2 B_1 \nonumber \\
		& & +\,\case{1}{2}m_\pi^2 [B_1 + B_2], 
	\label{eq:unquenchedBBd} 
\end{eqnarray}
where $G_\pm = (1\pm3g^2)/(4\pi f)^2$, $f=130$~MeV, 
and $g^2$ is the $B$-$B^*$-$\pi$ coupling.
$\chi$PT for the ${B_s}$ meson does not yield pion
contributions~$I(m_\pi^2)$.
A~rough guide for $g^2$ comes from the $D^*$ width.
Below I shall take $g^2=0.35\pm0.15$ to reflect the 
difference between the $D$ and $B$ systems.

The dynamics of the QCD scale is lumped into the constants $f_i$ and
$B_i$.
The long-distance radiation of pions (and $K$ and $\eta$ mesons) yields
$I(m^2)$, which, in detail, depends on how $\chi$PT is renormalized.
(The constants cancel the scheme dependence.)
In a mass-independent scheme
\begin{equation}
	I(m^2) = m^2 \ln(m^2/\mu^2).
	\label{eq:chilog}
\end{equation}
All schemes take this form when $m^2\ll\mu^2$; they differ when 
$m^2\sim\mu^2$.
The natural choice for $\mu$ is 500--700~MeV, but in most lattice 
calculations the ``pion'' mass is this large too.
The extrapolation then depends on how one treats~$I(m_\pi^2)$, 
and that is why chiral extrapolations are uncertain.

\begin{figure}[b!p]
	\vspace{-14pt}
	\includegraphics[width=75mm]{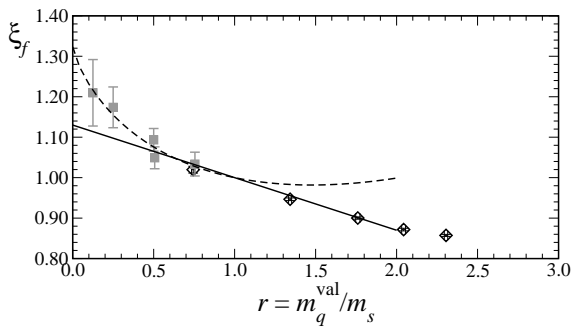}
	\vspace{-36pt}
	\caption[fig:xi]{$\xi_f$ vs.\ $r$ unquenched.
	Open symbols denote $n_f=2$ Wilson sea quarks \cite{Aoki:2003xb};
	solid gray symbols denote $n_f=2+1$ improved staggered 
	sea quarks~\cite{Wingate:2003ni}.
	The solid line (dashed curve) is JLQCD's linear fit (a chiral-log 
	fit~\cite{Kronfeld:2002ab}) to JLQCD's data.
	Statistical errors only.}\vspace{-4pt}
	\label{fig:xi}
\end{figure}

The problem with the chiral extrapolation is shown in Fig.~\ref{fig:xi}, 
which plots $\xi_f$ vs.\ 
$r=m_q^{\rm val}/m_s=m^2_{\rm PS}/(2m_K^2-m_\pi^2)$.
The black open points come from JLQCD's simulations with
$n_f=2$~\cite{Aoki:2003xb}, replotted from the original such that
$\xi_f=1$ when $r=1$.
These data are all at $r\ge0.7$, and we would like reach $r=0.04$.
The curves are two fit Ans\"atze: a linear fit~\cite{Aoki:2003xb},
yielding $\xi_f=1.13$; and a chiral-log fit, in which the slope is used 
to determine the constant $f_2$~\cite{Kronfeld:2002ab},
yielding $\xi_f=1.30$.
Both fits are extreme: the linear fit incorrectly omits the pion cloud,
the chiral-log fit trusts the form~(\ref{eq:chilog}) even when 
$m_{\rm PS}>500$~MeV,
where $\chi$PT is a poor description of hadrons~\cite{Sanz-Cillero:2003fq}.
Because these data alone do not verify the chiral form~(\ref{eq:chilog}),
the uncertainty in $\xi_f$ is large.

The gray points in Fig.~\ref{fig:xi} are preliminary results from HPQCD
\cite{Wingate:2003ni} on the MILC ensembles, with $n_f=2+1$ and
$r\le0.8$.
Although the statistical errors are still too large to get too excited,
it is striking how closely they track the chiral-log curve 
(which is fit to JLQCD only).
Taking just the smallest mass point suggests $\xi_f\gtrsim1.2$.

For the benefit of those interested in the CKM fit, the rest of this 
section recommends values for the mixing parameters.
Let us start with the bag factors, whose chiral logs are are multiplied
by $1-3g^2\approx-0.05$ and, thus, are small:
\begin{eqnarray}
	\hat{B}_{B_s} = 1.31 & \pm & 0.10 , \quad
	\hat{B}_{B_d} = 1.26 \pm 0.10 , \\
	\xi_B = 1.022 & \pm & 0.018 ,
\end{eqnarray}
symmetrizing JLQCD's ranges~\cite{Aoki:2003xb}.
These values are probably robust, because the bag factors seem to be 
insensitive to $n_f$~\cite{Yamada:2002wh}, and the chiral uncertainty is 
a fraction of the total.

Recommendations for $f_{B_s}$, $f_{B_d}$, and $\xi$ are less
straightforward.
$f_{B_s}$ should be gold-plated~\cite{Davies:2003ik,Gottlieb:2003lt},
but the comparison of JLQCD and HPQCD
\begin{eqnarray}
	f_{B_s} = 215(9)(14)~{\rm MeV} & \; (n_f=2) &
		 \hfill \cite{Aoki:2003xb} , \hfill \label{eq:fBsJLQCD} \\
	f_{B_s} = 260(7)(29)~{\rm MeV} & \; (n_f=2+1) &
		\hfill \cite{Wingate:2003ni} \hfill , \label{eq:fBsHPQCD}
\end{eqnarray}
is unsettling.
The first error bar is statistical; the second comes from systematics,
such as matching, that are partly common.
A further uncertainty in Eq.~(\ref{eq:fBsJLQCD}) comes from converting to~MeV
with the mass of the unstable $\rho$ meson.
\pagebreak
Converting with $\Upsilon$ splittings would increase $f_{B_s}$,
perhaps by 15\%~\cite{AliKhan:2001jg}.
Alas, Eq.~(\ref{eq:fBsHPQCD}) is still preliminary.
It is impossible to weigh these issues objectively.
My recommendation (for this year)~is
\begin{equation}
	f_{B_s} = 240 \pm 35~\textrm{MeV}.
\end{equation}
%
%
%

To estimate $\xi$, the ratio $\Xi=\xi_f\xi_Bf_\pi/f_K$ is useful.
The chiral log largely cancels between $\xi_f$ and
$f_\pi/f_K$ \cite{Becirevic:2002mh}, and the sensitivity to $g^2$
partly cancels between $\xi_f$ and $\xi_B$ \cite{Kronfeld:2002ab}.
From JLQCD's data~\cite{Aoki:2003xb}
I~obtain $\xi=(f_K/f_\pi)\Xi=1.23\pm0.06$;
HPQCD quotes (preliminarily) 1.22--1.34~\cite{Wingate:2003ni},
though most of the uncertainty here is statistical.
I shall quote round numbers
\begin{equation}
	\xi = 1.25 \pm 0.10
\end{equation}
with the fervent hope that next year's reviewer can quote a 
smaller, yet robust, error bar.

\section{SUMMARY AND PERSPECTIVE}
\label{sec:summary}

Most of this review has covered methodology.
It may be dull, but I hope that this focus will help us to meet the 
challenge of flavor physics.
The principal concerns---the quenched approximation, heavy-quark
discretization effects, and the chiral extrapolation---are easy enough 
to list.
Progress in unquenched 
calculations~\cite{Davies:2003ik,Gottlieb:2003lt} says that we must 
confront the other two problems.

Here, and elsewhere~\cite{Kronfeld:2002pi}, I have argued that we should
attack both problems by separating scales with effective field
theories.
This gives us a framework that is theoretically sound, and familiar to 
other theorists, and experimenters too.
As long as we are not naive in applying effective field theories, we 
have good reason to believe that our error bars will be robust and 
persuasive.

Finally, although it is clear that lattice QCD has solid foundations, 
let us remember that we carry out numerical simulations.
These are not easy for outsiders (even other experts) to grasp fully.
It therefore always helps to have tests of the whole apparatus.
In the case of heavy quarks, a good set of checks come from quarkonium,
where the spectrum  and also many electromagnetic decay amplitudes are
well measured and (for us) gold-plated.
Even better for $B$ physics are upcoming measurements in $D$ physics.
CLEO-$c$ will soon measure leptonic and semi-leptonic \pagebreak
$D$ decays to a few percent. 
Assuming CKM unitarity, one then has a measurement of decay constants 
and form factors.
If we can come to grips with the chiral extrapolation,
lattice QCD has a chance to predict the results.
Let's not squander it.

\vspace{0.5em}
I thank
C. Davies,
A. Gray,
S. Hashimoto,
P. Lepage, 
P. Mackenzie,
C. Maynard,
T. Onogi,
S. Sharpe,
J. Shigemitsu,
M. Wingate, and
N. Yamada
for useful discussions.
Fermilab in operated by Universities Research Association Inc., under
contract with the U.S.\ Department of Energy.

\newcommand{\collab}[1]{\ [#1]}

\end{document}